\documentclass[letter,longauth,traditabstract]{aa}  

\usepackage{graphicx}
\usepackage{txfonts}
\usepackage{color}
\usepackage{soul}
\usepackage{placeins}

\def\mstar  {$M_{\star}$}
\def\macc   {$\dot{M}_{\rm acc}$}
\def\lacc   {$L_{\rm acc}$}

\def\mdisk {$M_{\rm disk}$}

\def\msun {$M_{\odot}$}
\def\lsun {$L_{\odot}$}
\def\lstar {$L_\star$}

\newcommand{\teff}{$T_{\rm eff}$}

\usepackage{xcolor}

\begin{document}

   \title{PENELLOPE III. The peculiar accretion variability of XX Cha and its impact on the observed spread of accretion rates \thanks{Based on observations collected at the European Southern Observatory under ESO programmes 084.C-1095 and 106.20Z8.}}

   \author{R.A.B. Claes 
          \inst{\ref{instESO}}
          \and
          C.F. Manara\inst{\ref{instESO}}
          \and
          R. Garcia-Lopez\inst{\ref{instUCD},\ref{instDIAS}}
          \and
          A. Natta\inst{\ref{instDIAS}}
          \and 
          M. Fang \inst{\ref{instMFang}}
          \and
          Z. P. Fockter \inst{\ref{Akop1},\ref{ZPF2}}
          \and
          P. \'Abrah\'am \inst{\ref{Akop1},\ref{ZPF2}}
          \and
          J.M. Alcal\'{a}\inst{\ref{instNA}}
         \and 
          J. Campbell-White \inst{\ref{Justyn}}
          \and
          A. Caratti o Garatti \inst{\ref{instUCD},\ref{instDIAS},\ref{instNA}}
          \and 
          E. Covino\inst{\ref{instNA}}    
          \and
          D. Fedele\inst{\ref{instTO},\ref{instFI}} 
          \and
          A. Frasca\inst{\ref{AFrasca}}
          \and
          J.F. Gameiro \inst{\ref{JFG1},\ref{JFG2}}
          \and
          G.J. Herczeg\inst{\ref{KIAA},\ref{GJH} }
          \and
          \'A. K\'osp\'al \inst{\ref{Akop1},\ref{Akop2},\ref{ZPF2}}
          \and 
          M. G. Petr-Gotzens\inst{\ref{instESO}}
           \and
          G. Rosotti \inst{\ref{GRossoti}}
         \and 
          L. Venuti \inst{\ref{instLvenuti}}
          \and 
          G. Zsidi\inst{\ref{Akop1},\ref{ZPF2}}
           } 

  \institute{European Southern Observatory, Karl-Schwarzschild-Strasse 2, 85748 Garching bei M\"unchen, Germany\label{instESO}\\
              \email{Rik.Claes@eso.org}
         \and
 School of Physics, University College Dublin, Belfield, Dublin 4, Ireland\label{instUCD}
\and 
           School of Cosmic Physics, Dublin Institute for Advanced Studies, 31 Fitzwilliam Place, Dublin 2, Ireland\label{instDIAS}
\and
Purple Mountain Observatory, Chinese Academy of Sciences, 10 Yuanhua Road, Nanjing 210023, People's Republic of China\label{instMFang}
\and
Konkoly Observatory, Research Centre for Astronomy and Earth Sciences, E\"otv\"os Lor\'and Research Network (ELKH), Konkoly-Thege Mikl\'os \'ut 15-17, 1121 Budapest, Hungary \label{Akop1}
\and
ELTE E\"otv\"os Lor\'and University, Institute of Physics, P\'azm\'any P\'eter s\'et\'any 1/A, 1117 Budapest, Hungary \label{ZPF2}
\and
INAF -- Osservatorio Astronomico di Capodimonte, via Moiariello 16, 80131 Napoli, Italy\label{instNA}
\and 
SUPA, school of science and engineering, university of dundee, nethergate, dundee dd1 4hn, U.K.\label{Justyn}
\and
INAF -- Osservatorio Astrofisico di Torino, Via Osservatorio 20, I-10025 Pino Torinese, Italy\label{instTO}
\and 
INAF -- Osservatorio Astrofisico di Arcetri, L.go E. Fermi 5, 50125 Firenze, Italy\label{instFI}
\and
INAF - Osservatorio Astrofisico di Catania, via S. Sofia, 78, 95123 Catania, Italy\label{AFrasca}
\and
Departamento de F\'isica e Astronomia, Faculdade de Ci\^encias, Universidade do Porto, Rua do Campo Alegre 687, PT4169-007 Porto, Portugal \label{JFG1}
\and
Instituto de Astrof\'isica e Ci\^encias do Espa\c{c}o, Universidade do Porto, CAUP, Rua das Estrelas, PT4150-762 Porto, Portugal \label{JFG2}
\and 
Kavli Institute for Astronomy and Astrophysics, Peking University, Yiheyuan 5, Haidian Qu, 100871 Beijing, China \label{KIAA}
\and
Department of Astronomy, Peking University, Yiheyuan 5, Haidian Qu, 100871 Beijing, China\label{GJH}
\and
Max Planck Institute for Astronomy, K\"onigstuhl 17, D-69117 Heidelberg, Germany \label{Akop2}
\and
School of Physics and Astronomy, University of Leicester, Leicester LE1 7RH, UK\label{GRossoti}
\and
SETI Institute, 339 Bernardo Avenue, Suite 200, Mountain View, CA 94043, USA\label{instLvenuti}
}

   \date{Received May 26, 2022; accepted -}

\titlerunning{The peculiar accretion variability of XX Cha}

 
  \abstract
 {The processes regulating protoplanetary disk evolution are constrained by studying how mass accretion rates scale with stellar and disk properties. The spread in these relations can be used as a constraint to the models of disk evolution, but only if the impact of accretion variability is correctly accounted for. While the effect of variability might be substantial in the embedded phases of star formation, it is often considered limited at later stages. Here we report on the observed large variation in the accretion rate for one target, XX Cha, and we discuss the impact on population studies of classical T Tauri stars. The mass accretion rate determined by fitting the UV-to-near-infrared spectrum in recent X-Shooter observations is compared with the one measured with the same instrument 11 years before. XX Cha displays an accretion variability of almost 2 dex between 2010 and 2021. Although the timescales on which this variability happens are uncertain, 
 XX Cha displays an extreme accretion variability for a classical T Tauri star. If such behavior is common among classical T Tauri stars, possibly on longer timescales than previously probed, it could be relevant for discussing the disk evolution models constrained by the observed spread in accretion rates. 
 Finally, we remark that previous studies of accretion variability based on spectral lines may have underestimated the variability of some targets.}


   \maketitle
%

\section{Introduction}

While planets are forming, the protoplanetary disks they are born in evolve under the effect of several processes. Of particular relevance is how the material is transported through the disk and how it is accreted onto the central star \citep{hartmann16}. Accretion regulates the final stellar mass, and it is commonly used in combination with the stellar and disk mass to constrain the models driving the global evolution of disks \citep{manara22}.
  
The accretion process is intrinsically highly variable \citep[e.g.,][]{stauffer14} on timescales of minutes to years \citep[e.g.,][]{Fang2013,costigan14}, with different magnitudes of variability \citep[e.g.,][]{hillenbrand15,Fischer22}. It has often been questioned whether accretion variability can explain the spread of $\sim$2-3 dex observed in the relations between accretion rates and stellar or disk masses \citep[e.g.,][]{manara20,manara22}. If not dominated by accretion variability, this spread is a way to test and constrain disk evolution models, in particular to highlight the limits of the viscous evolution scenario \citep[e.g.,][]{mulders17,lodato17,manara20,manara22}. Independently from each other, several works have shown that, typically, accretion variability peaks at $\lesssim$0.5 dex on timescales of a few weeks \citep{costigan12,costigan14,biazzo12,biazzo14,venuti14,frasca15,zsidi22}. Such variability is too small to explain the observed scatter. However, little data exist probing longer timescales of decades, and this is usually limited to single band photometric studies, which would hardly distinguish between variations in the stellar photosphere or accretion variability. Finally, a tiny fraction of young stars, named FU Orionis and EXor, have been identified to show a strong ($\sim$2-3 dex) increase in their accretion rates lasting years \citep[e.g.,][]{audard14,Fischer22}. Since they are rare, they are not considered to have a significant impact on the observed relations.

Here we present and analyze the case of the highly variable young stellar object (YSO) XX Cha, observed with the same spectrograph in 2010 \citep{manara16a,manara17a} 
and then in 2021. We then discuss the implications on studies of disk evolution based on accretion and stellar or disk masses.

\begin{figure*}[th!]
    \centering
    \includegraphics[width=0.85\textwidth]{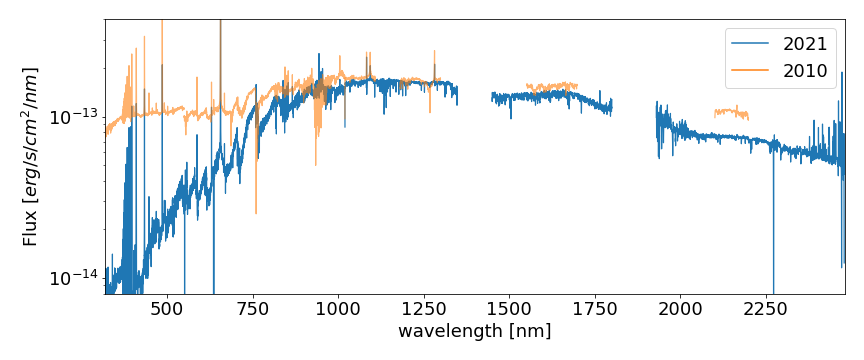}
    \caption{Comparison between the spectra of XX Cha as observed with X-Shooter in January 2010 (orange) and June 2021 (blue). For the sake of clarity of the figure, the spectra were convolved by a Gaussian with a standard deviation of three times the spectral resolution. }
    \label{fig:xs_old_new}
\end{figure*}

\section{Target, observations, and data reduction}

The target of this study is XX Cha (2MASS J11113965-7620152, Ass Cha T 2-49), a classical T Tauri star (CTTS) in the Chamaeleon I region, at a distance of 192 pc \citep{Gaia2021}. This star is part of a very wide binary system with CHX 18N \citep{Kraus2007}, with an angular separation of 24.38\arcsec$\sim$4,600 au. 

XX Cha hosts a disk that was detected with Atacama Large Millimeter Array (ALMA) in the Band 7 continuum \citep{pascucci16}, but it was not detected in the $^{13}$CO gas emission \citep{long17}. The corresponding disk dust mass is 8.12 $M_\oplus$ \citep{manara22}.

XX Cha is known to have a significant mid-infrared variability. \citet{kospal12} compared the ISO/ISOPHOT-S and Spitzer IRS spectra of this target, and found a strong decadal change in the 5-12 $\mu$m range, with a peculiar wavelength dependence. XX Cha is one of the only sources where an anti-correlation was found in the silicate emission feature at $~10\mu$m with the continuum flux between $6-8\mu$m, as opposed to a correlation. Such a variability may be caused by variable shadowing of the silicate emitting region \citep{kospal12}.

\subsection{Spectroscopic observations}

XX Cha was observed from the ESO Very Large Telescope (VLT) with the X-Shooter instrument. It is a medium-resolution spectrograph simultaneously observing in three wavelength ranges, called the UV-Blue (UVB) ($\sim$300–560 nm), Visible (VIS) ($\sim$560–1024 nm), and Near-IR (NIR) ($\sim$1020–2480 nm) arms \citep{vernet11}. 

The first observation was taken on 19 January 2010 (UT 07:39:06) in visitor mode, and it was analyzed by \citet{manara16a}. The target was observed a second time on 5 June 2021 (UT 03:45:18) as a part of the PENELLOPE Large Program \citep{manara21}. The latter observation is presented here for the first time. Both observations used a combination of a short exposure with a wide set of slits - 5.0\arcsec\, wide, except for the NIR arm in 2010 when a 1.5\arcsec\, slit was used to measure absolute fluxes, followed by a long nod-on-slit integration with a narrower set of slits - 1.0\arcsec-0.4\arcsec-0.4\arcsec\, in the UVB, VIS, and NIR arms, so as to achieve a spectral resolution $R\sim$ 5400, 18400, and 11600 in the three arms, respectively.

The reduction of the spectra taken was performed using the ESO X-Shooter pipeline \citep{xspipe} v1.3.2 and v3.5.0 for the 2010 and 2021 epochs, respectively. The final flux calibration was performed by rescaling the narrow slit spectrum to the wide slit one as described by \citet{manara16a,manara21}.
The flux calibration accuracy for the 2010 epoch of XX Cha was found by \citet{Rugel2018} to be $\sim 4\%$. We expect a similar value for the 2021 epoch.
In 2010, the low signal to noise of the flux standard star in the NIR arm did not allow for a proper flux calibration of most of this arm, and telluric correction could also not be performed. Telluric lines were  removed from the VIS and NIR parts of the 2021 spectrum using the molecfit tool \citep{molecfit1}.

\subsection{Photometric data}\label{sec:phot_data}
The American Association of Variable Star Observers (AAVSO) obtained photometric data for XX~Cha to complement the TESS and Hubble Space Telescope (HST) ULLYSES \citep{ullysesDR1,espaillat22} observations\footnote{\parbox{5.5cm}{\url{https://www.aavso.org/hst-ulysses southern-t-tauri-campaign-01}}} from April to August 2021. We collected this dataset, covering $B$, $V$, $R$, and $I$ band photometry, from the AAVSO website \footnote{ \url{https://www.aavso.org/data-download}} \citep{Kafka2020}.

Optical photometry in the $V$, $R$, and $I$ band is also available for periods spanning from April 2010 to May 2010 and January 2013. 
The 2010 photometry contains ten observations obtained with the REM telescope (Pr.Id. 21703, PI: Michel Curé). The 2013 photometry contains four epochs obtained with the ANDICAM instrument on the CTIO 1.3m telescope (Project ID: ESA-12B-0001, PI: Ágnes Kóspál).
The light curve and additional information is reported in Appendix \ref{photometry}.

\section{Analysis}

\subsection{Continuum variations}\label{sec:cont}
A comparison between the two X-Shooter spectra can be seen in Fig.~\ref{fig:xs_old_new}. 
The spectrum obtained in 2010 is significantly brighter than the one taken in 2021 at wavelengths shorter than $\sim$800 nm, suggestive of either a higher accretion rate in 2010 or different extinction. 
The two spectra look quite similar at longer wavelengths; although, the $K$-band spectrum in the 2010 epoch is, on average, $\sim 1.4$ times brighter than in the 2021 epoch, hinting at a difference in the inner disk properties.

Synthetic photometry was performed on both X-Shooter spectra using the PYPHOT package\footnote{ \url{https://mfouesneau.github.io/docs/pyphot/index.html}} in the Johnson-Cousins $B$, $V$, $R$ and $I$ bands. 
The obtained values are listed in Table~\ref{SynthPhot}. A good agreement between the 2021 spectrum and almost simultaneous photometry can be seen in Fig.~\ref{fig:synthPhot}. The 2021 X-Shooter observation was obtained when the star was at a minimum of the short-term light curve. No simultaneous photometry is available for the 2010 epoch. However, the synthetic photometry of the 2010 spectrum appears to fall close to the maximum brightness  observed in the available photometry. We conclude that the X-Shooter observations are taken close to the extremes of the known photometric variability of XX Cha.

Variations in the circumstellar extinction can explain dippers and dimming events of peculiar stars \citep[e.g.,][]{Schisano2009,bouvier13,facchini16,schneider18,koutoulaki19}. To exclude this possibility, Fig.~\ref{fig:VVSv-r}-\ref{fig:bVSb-v} show the color-magnitude diagram of XX Cha, and they indicate that the variations between the two epochs are incompatible with just extinction variations. 
Similarly, flux ratios between the two spectra are compared with typical reddening laws for increasing values of $A_V$ and $R_V$ in Fig. \ref{Fig:FluxRatioVSred}, showing that the observed variations cannot be fully explained by a difference in extinction. 

Finally, the lithium line at 670.78 nm is found to have a lower equivalent width in 2010 $(\sim 0.150)$ when compared to 2021 $(\sim 0.550)$ (Fig. \ref{fig:LiLine}). This is due to an increase in the veiling in 2010, which is related to an increase in the accretion rate. The analysis of the spectra with ROTFIT \citep{frasca2017} indeed leads to a measured veiling at 620 nm of 6.4 in 2010 and of 0.6 in 2021.

\subsection{Stellar and accretion properties}

The stellar and accretion parameters were derived by fitting the UV excess in the X-Shooter spectra with the procedure described by \citet{manara13b} (see Appendix~\ref{App:fit}). The 2010 data were analyzed with this method by \citet{manara16a}. Here, for 2010, we use the values reported by \citet{manara22}, which were computed by rescaling the distance to 192 pc, assuming the relation between spectral type and temperature by \citet{HH14}, and the non-magnetic evolutionary tracks by \citet{Feiden2016} to obtain the stellar mass. These values are reported in Table~\ref{sec2:Param}. 

With the same method and the same assumptions, we fit the 2021 spectrum, and we derived a spectral type consistent with the 2010 result within half of a subclass, but we obtain significant differences in the other parameters. In particular, the mass accretion rate (\macc) is found to be smaller by $\sim$2 dex than in 2010, and $A_V$ is found to be smaller by 0.7 mag.
Since veiling can make the determination of the stellar parameters more uncertain \citep[e.g.,][]{calvet98}, we repeated the fit of the 2010 spectrum, using the same photospheric template and extinction as derived in the 2021 epoch. 
We obtain a lower \macc\, than previously reported (see  Table \ref{sec2:Param}, column 2010*), leading to a difference in \macc\, of $\sim 1.4$ dex compared to the 2021 observations.

\begin{table}[t]
\centering
\caption{Stellar and accretion parameters obtained for both epochs}
\label{sec2:Param}
\begin{tabular}{c|ccc}
\hline \hline 
 & \multicolumn{3}{c}{Epoch of observation} \\ 
Property & 2010 & 2021 & 2010* \\ \hline 
SpT & M3.5 &   M3  & M3* \\
\teff~ (K) & 3300  &    3410 & 3410* \\
$A_v$ (mag)& 1.0 &   0.3  & 0.3* \\
\lstar~ (\lsun)& 0.42 &  0.30&  0.39 \\
$\log$ \lacc (\lsun)& -0.65 &   -2.34 & -1.06 \\
\mstar~ (\msun) & 0.24 &   0.30 & 0.30\\
 $\log$ \macc (\msun /yr) & -7.14 & -9.06 & -7.69\\
 \hline
\end{tabular}
\tablefoot{The values listed under ``2010*'' report the results obtained fitting the 2010 epoch with the same photospheric template and $A_V$ as the values obtained fitting the 2021 epoch. Typical uncertainties are the following: 0.1 for  $A_v$ , 0.2 dex for \lstar, 0.1 dex for \mstar,  0.25 for \lacc\  \citep{manara17a}, and 0.35 for \macc \ \citep{manara22}. } 
\end{table}

\subsection{Accretion properties from the emission lines}\label{sec:lline}

Starting from the relations between the line luminosity ($L_{\rm line}$) and the accretion luminosity (\lacc) by \citet{alcala17}, we calculated \lacc\, using different emission lines and assuming $A_V = 0.3$ mag for both observations.  
The mean values of \lacc\, measured from the line luminosity agree within the uncertainties with the corresponding value from the UV-excess fitting at each epoch (Fig.~\ref{fig:LaccLines2010*} for 2010*, $\log(\langle L_{\rm acc,lines}/L_\odot\rangle )= -1.50 \pm 0.35 $; Fig.~\ref{fig:LaccLines2021} for 2021, $ \log(\langle L_{\rm acc,lines}/L_\odot\rangle )= -1.98 \pm 0.35$).
On the other hand, the difference between the two values of \lacc\, from the line luminosity is smaller than when computed from the UV excess.

We note that the line profiles are very different between the two epochs (Fig.~\ref{fig:balmer}-\ref{fig:CaKLine}). In particular, the 2010 observations show a stronger red-shifted absorption and a wider blueshifted emission in the hydrogen emission lines. Both aspects are in line with the stronger accretion rate measured in 2010. 

We also considered additional estimates of \lacc\, from line emission available from an observation of XX Cha in 2009 by \citet{Antoniucci2011}. We scaled the luminosity of the H$\alpha$,  Pa$\beta$, and Br$\gamma$ lines, accounting for the different distance and extinction adopted here. We find an average accretion luminosity of $ \log(\langle L_{\rm acc,lines}/L_\odot\rangle) = -1.3 \pm 0.5$ dex, which is slightly higher but compatible with the one measured in 2010 from the line luminosity. We derived an accretion rate of $\log$ \macc (\msun /yr)$=-8.0 \pm 0.7$ dex, assuming the stellar mass obtained for the 2010* epoch. This value is consistent with the accretion properties found in 2010*.

\section{Discussion}

\subsection{Variability of XX Cha in context} 
The $\gtrsim$1.4 dex variations in accretion rates measured in XX Cha poses the question of whether this target could be part of either the FU Orionis or EXor variable classes \citep[e.g.,][]{Fischer22}. 
We exclude the former, since FU Orionis stars experience sudden jumps in accretion rate to about \macc$\sim 10^{-4} M_\odot$/yr \citep{kospal2011FUor}  which last for timescale of decades to centuries. 
Both the value of \macc\, and the timescales of the photometric evolution (Fig.~\ref{fig:synthPhot}) are very different than typical FU Orionis outbursts.
During an outburst, several spectral features in the spectra of FU Orionis stars are the opposite of what we observe in XX Cha, for example in FU Orionis stars the Pa$\beta$ line is observed in absorption and the CaII infrared triplet is not detected \citep{Connelley2018}. 

EXor stars, on the other hand, display an increase in \macc\, of $\sim1 - 2$ dex, similar to XX Cha. These variations occur in an outburst that typically lasts several months to a year.
In EXor-type outbursts, a large number of emission lines can be detected in the near-infrared \citep[e.g.,][]{kospal11EXor}. Such lines are not evident in the 2010 spectrum of XX Cha, but \citet{Antoniucci2011} in 2009 observed ro-vibrational CO transitions
at $\lambda>2300$ nm in emission, which is a typical feature of EXor outbursts \citep{Fischer22}. As discussed in Sect.~\ref{sec:lline}, XX Cha was accreting at a similar \macc\,, both in our 2010 observations and in 2009.
Both the high accretion luminosity and CO emission found by \citet{Antoniucci2011} suggest that XX Cha may have experienced an EXor-type outburst in 2009, with our 2010  observations being taken at a later stage during this outburst.
However, the lack of data between the two epochs does not allow us to confirm or exclude this possibility.

\begin{figure}
    \centering
    \includegraphics[width = 0.49\textwidth]{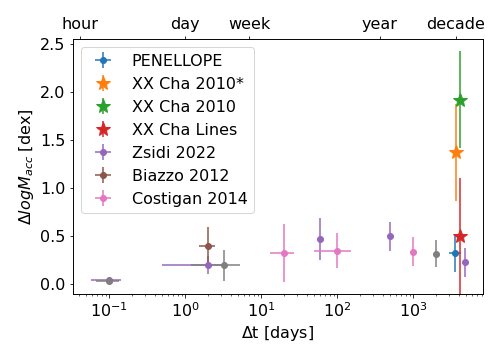}
    \caption{Relative accretion rate variations as a function of the time difference between observations. The points indicate the mean variability and the errorbars mark their standard deviations. We note that literature values, with the exeption of those from \citet{zsidi22}, are drawn from measurements in several stars.}
    \label{fig:dMaccVsDt}
\end{figure}

Even if it is unclear whether XX Cha is an EXor variable star, it is instructive to compare its extreme accretion variability to previous samples of CTTS. We show in Fig.~\ref{fig:dMaccVsDt} the extent of its variability in comparison with literature results \citep{biazzo12,costigan14,zsidi22} and with the preliminary analysis of 11 other targets from the PENELLOPE survey \citep[][Garcia-Lopez et al. in prep.]{manara21}. 
The data for XX Cha are displayed using the interval between the observations, and both of the values measured from UV excess and line luminosity. Based on our data, we cannot distinguish between a timescale for the observed variability of 11 years or of $\sim 22$ days, an indication of which could be determined in reviewing the observed light curve in 2021 (Fig.~\ref{fig:synthPhot}).
Either way, the accretion variability of XX Cha measured from the UV excess is more extreme than literature results for CTTS, including the similarly analyzed PENELLOPE targets. Among other possible objects showing a potentially similar variability as XX Cha, we can cite V347 Aur, since it has periodic $V$ band variations with an amplitude of $\sim 2$ mag on timescales $\sim 160$ days. However, other effects, such as a variable extinction, could significantly contribute to the photometric variability.

The accretion variability measured from the line luminosity ($\sim$0.5 dex) is instead in line with literature results. We strongly advocate that studies of accretion variability should be performed with flux-calibrated broad wavelength range spectra.
A possible explanation for the different results obtained from the UV excess and emission lines could be a partial occultation due to an inner disk warp. In this case, similar to AA-Tau \citep{bouvier13, Schneider2015}, the UV radiation from the shock region on the stellar surface is obscured by the disk, whereas the line emission coming from the accretion columns would be less obscured. In turn, this would result in comparatively higher line emission in the epoch with lower UV excess, and vice versa. If such a differential extinction could explain the discrepancy between a line and UV excess, this would affect the measurements of \macc\, in population studies based on a single value of $A_V$ and on UV-excess measurements. Therefore, when we discuss the impact of this accretion variability on the observed spreads in the next section, the difference measured from the UV excess is the relevant one.

\subsection{Impact on accretion rates spread}

\begin{figure}[t!]
   \centering
   \includegraphics[width=0.49\textwidth]{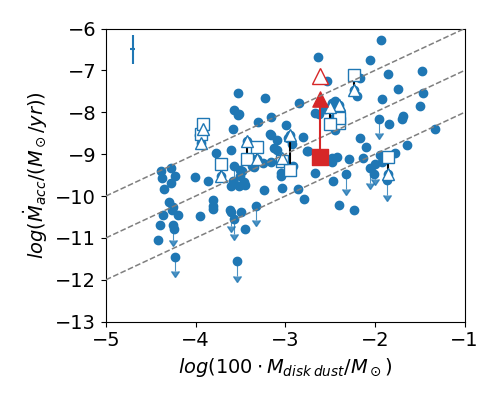}
      \caption{Measured \macc \, and \mdisk\, for targets in the Lupus and Chamaeleon~I star-forming regions (blue filled dots). The red filled (empty) triangle and square indicate the 2010* (2010) and 2021 mass accretion rates of XX Cha, respectively.
      The dashed lines display the \mdisk/\macc\: ratio of 0.1, 1, and 10 Myr, respectively. 
      The white empty squares and triangles indicate targets observed in 2021 in the PENELLOPE program \citep{manara21} and around 2010, respectively. 
              }
         \label{fig::macc_mdisk}
   \end{figure}

The observed distributions on the \macc\: - \mdisk\: and \macc\: - \mstar\: planes are a useful test bench of the mechanisms driving protoplanetary disk evolution \citep{manara22}. To understand how a significantly variable object such as XX Cha impacts these relations, we plotted in Fig.~\ref{fig::macc_mdisk} and Fig.~\ref{fig:macc_mstar} the data taken from \citet{manara22} for the Chamaeleon I and Lupus star-forming regions, as well as the two measured values for XX~Cha.  
This target does not appear to be an outlier in the distribution of \macc\, values in either 2010 or 2021, since its variability is still smaller than the observed spread, and the measured values of \macc\, are well in line with typical measurements for similarly massive stars or disks. In this specific case, the measured variability therefore has no impact on the observed spread. The effect would be significant if a larger number of targets would undergo similarly significant variability, or even a stronger one. 

Based on currently available information, XX Cha seems still to be a rare outlier. Indeed, the preliminary analysis of other PENELLOPE data points to the fact that, even looking at the UV excess, typical variability is $\lesssim$0.5 dex (see both Fig.~\ref{fig:dMaccVsDt} and \ref{fig::macc_mdisk}). Also literature exploration of both short- and long-term variability seems to converge on this result. \citet{venuti14,venuti15} used photometrically derived UV-excess measurements to determine that, overall, variations on UV excess over timescales of years are statistically consistent with those on timescales lasting for weeks, with a typical variation of 0.4 $\pm$ 0.3 dex when considering all CTTS, and 0.6$\pm$0.5 dex when considering only CTTS with \macc\, variations $>$1 dex over $\sim$2 week timescales. Only about 14\% of stars in their sample show additional long-term variations in UV excess beyond 0.5 dex.
If these results are confirmed, we can assume that the majority of the spread of accretion rates is intrinsic, and not related to accretion variability. 

However, we cannot base our estimates purely on line equivalent width or luminosity estimates since we have shown that these could lead to an underestimation of the extent of the variability. Further monitoring with broad wavelength range spectra is needed to constrain the fraction of sources showing large variability, so as to better constrain the effect of accretion variability on the observed relations, making them a more powerful tool to study protoplanetary disk evolution.   

Finally, we note that the large variability in the accretion rate of XX Cha would have an effect on the measured value of $t_{\rm acc}$=\mdisk/\macc\ anyways. This quantity is used, for example, to discriminate between targets whose evolution is regulated by viscous evolution \citep[e.g.,][]{lodato17,mulders17}, magnetohydrodynamic wind-driven evolution \citep[e.g.,][]{mulders17,tabone21}, or even by other internal or external disk evolution processes \citep[e.g.,][]{rosotti17}. While population studies are unaffected, since the spread is not sensitive to this magnitude of variability, results on individual targets must consider this additional cause of uncertainty due to variability before firmly assessing whether the displacement of a target on the \macc-\mdisk\, plane is due to one or another disk evolution mechanism. The interpretation of individual values should be confirmed with multiple epochs of observation, when possible.

\section{Conclusions}
We report a variability in accretion rates for XX Cha of $\gtrsim$1.4 dex between 2010 and 2021, when measured from the UV excess. This change in accretion rate is much larger than the typical accretion variability of $\lesssim$0.5 dex found in other CTTS. 
However, a smaller $\sim$0.5 dex variability is obtained for XX Cha when measured from the luminosity of emission lines.
This result highlights the need to monitor accretion using flux-calibrated broad wavelength range spectra.
Although large, the measured variability of XX Cha has no impact on the observed spread of \macc\, in the relations with \mdisk\, or \mstar. Since the variability of XX Cha seems rare to date, this confirms that the spread of \macc\, is related to disk evolution. However, we suggest that more extensive exploration of accretion variability on longer timescales than previously probed must be performed, so as to firmly asses its importance.


\begin{acknowledgements}

We thank M. Benisty and \textbf{J. Bouvier} for discussions on this target.
R. Claes acknowledges the PhD fellowship of the International Max-Planck-Research School (IMPRS) funded by ESO. 
This work benefited from discussions with the ODYSSEUS team (HST AR-16129), \url{https://sites.bu.edu/odysseus/}.
Funded by the European Union under the European Union’s Horizon Europe Research \& Innovation Programme 101039452 (WANDA). Views and opinions expressed are however those of the author(s) only and do not necessarily reflect those of the European Union or the European Research Council. Neither the European Union nor the granting authority can be held responsible for them.
This project has received funding from the European Union's Horizon 2020 research and innovation programme under grant agreement No 716155 (SACCRED) and under the Marie Sklodowska-Curie grant agreement No 823823 (DUSTBUSTERS).
This work was partly funded by the Deutsche Forschungsgemeinschaft (DFG, German Research Foundation) - 325594231. 
This research received financial support from the project PRIN-INAF 2019 "Spectroscopically Tracing the Disk Dispersal Evolution".

R.G.Lopez acknowledges support by Science Foundation Ireland under Grant no. 18/SIRG/5597.

J.F.Gameiro was supported by fundacao para a Ci\^encia e Tecnologia (FCT) through the research grants UIDB/04434/2020 and UIDP/04434/2020.

G. Rosotti acknowledges support from an STFC Ernest Rutherford Fellowship (grant number ST/T003855/1)

\end{acknowledgements}

%
%

\bibliography{bibliography.bib}

\begin{thebibliography}{52}
\expandafter\ifx\csname natexlab\endcsname\relax\def\natexlab#1{#1}\fi

\bibitem[{{Alcal{\'a}} {et~al.}(2017){Alcal{\'a}}, {Manara}, {Natta}, {Frasca},
  {Testi}, {Nisini}, {Stelzer}, {Williams}, {Antoniucci}, {Biazzo}, {Covino},
  {Esposito}, {Getman}, \& {Rigliaco}}]{alcala17}
{Alcal{\'a}}, J.~M., {Manara}, C.~F., {Natta}, A., {et~al.} 2017, \aap, 600,
  A20

\bibitem[{{Antoniucci} {et~al.}(2011){Antoniucci}, {Garc{\'\i}a L{\'o}pez},
  {Nisini}, {Giannini}, {Lorenzetti}, {Eisl{\"o}ffel}, {Bacciotti}, {Cabrit},
  {Caratti o Garatti}, {Dougados}, \& {Ray}}]{Antoniucci2011}
{Antoniucci}, S., {Garc{\'\i}a L{\'o}pez}, R., {Nisini}, B., {et~al.} 2011,
  \aap, 534, A32

\bibitem[{{Audard} {et~al.}(2014){Audard}, {{\'A}brah{\'a}m}, {Dunham},
  {Green}, {Grosso}, {Hamaguchi}, {Kastner}, {K{\'o}sp{\'a}l}, {Lodato},
  {Romanova}, {Skinner}, {Vorobyov}, \& {Zhu}}]{audard14}
{Audard}, M., {{\'A}brah{\'a}m}, P., {Dunham}, M.~M., {et~al.} 2014, in
  Protostars and Planets VI, ed. H.~{Beuther}, R.~S. {Klessen}, C.~P.
  {Dullemond}, \& T.~{Henning}, 387

\bibitem[{{Biazzo} {et~al.}(2012){Biazzo}, {Alcal{\'a}}, {Covino}, {Frasca},
  {Getman}, \& {Spezzi}}]{biazzo12}
{Biazzo}, K., {Alcal{\'a}}, J.~M., {Covino}, E., {et~al.} 2012, \aap, 547, A104

\bibitem[{{Biazzo} {et~al.}(2014){Biazzo}, {Alcal{\'a}}, {Frasca}, {Zusi},
  {Getman}, {Covino}, \& {Gandolfi}}]{biazzo14}
{Biazzo}, K., {Alcal{\'a}}, J.~M., {Frasca}, A., {et~al.} 2014, \aap, 572, A84

\bibitem[{{Bouvier} {et~al.}(2013){Bouvier}, {Grankin}, {Ellerbroek}, {Bouy},
  \& {Barrado}}]{bouvier13}
{Bouvier}, J., {Grankin}, K., {Ellerbroek}, L.~E., {Bouy}, H., \& {Barrado}, D.
  2013, \aap, 557, A77

\bibitem[{{Calvet} \& {Gullbring}(1998)}]{calvet98}
{Calvet}, N. \& {Gullbring}, E. 1998, \apj, 509, 802

\bibitem[{{Cardelli} {et~al.}(1989){Cardelli}, {Clayton}, \&
  {Mathis}}]{cardelli98}
{Cardelli}, J.~A., {Clayton}, G.~C., \& {Mathis}, J.~S. 1989, \apj, 345, 245

\bibitem[{{Connelley} \& {Reipurth}(2018)}]{Connelley2018}
{Connelley}, M.~S. \& {Reipurth}, B. 2018, \apj, 861, 145

\bibitem[{{Costigan} {et~al.}(2012){Costigan}, {Scholz}, {Stelzer}, {Ray},
  {Vink}, \& {Mohanty}}]{costigan12}
{Costigan}, G., {Scholz}, A., {Stelzer}, B., {et~al.} 2012, \mnras, 427, 1344

\bibitem[{{Costigan} {et~al.}(2014){Costigan}, {Vink}, {Scholz}, {Ray}, \&
  {Testi}}]{costigan14}
{Costigan}, G., {Vink}, J.~S., {Scholz}, A., {Ray}, T., \& {Testi}, L. 2014,
  \mnras, 440, 3444

\bibitem[{{Espaillat} {et~al.}(2022){Espaillat}, {Herczeg}, {Thanathibodee},
  {Pittman}, {Calvet}, {Arulanantham}, {France}, {Serna}, {Hern{\'a}ndez},
  {K{\'o}sp{\'a}l}, {Walter}, {Frasca}, {Fischer}, {Johns-Krull}, {Schneider},
  {Robinson}, {Edwards}, {{\'A}brah{\'a}m}, {Fang}, {Erkal}, {Manara},
  {Alcal{\'a}}, {Alecian}, {Alexander}, {Alonso-Santiago}, {Antoniucci},
  {Ardila}, {Banzatti}, {Benisty}, {Bergin}, {Biazzo}, {Brice{\~n}o},
  {Campbell-White}, {Cleeves}, {Coffey}, {Eisl{\"o}ffel}, {Facchini}, {Fedele},
  {Fiorellino}, {Froebrich}, {Gangi}, {Giannini}, {Grankin}, {G{\"u}nther},
  {Guo}, {Hartmann}, {Hillenbrand}, {Hinton}, {Kastner}, {Koen}, {Mauc{\'o}},
  {Mendigut{\'\i}a}, {Nisini}, {Panwar}, {Principe}, {Robberto},
  {Sicilia-Aguilar}, {Valenti}, {Wendeborn}, {Williams}, {Xu}, \&
  {Yadav}}]{espaillat22}
{Espaillat}, C.~C., {Herczeg}, G.~J., {Thanathibodee}, T., {et~al.} 2022, \aj,
  163, 114

\bibitem[{{Facchini} {et~al.}(2016){Facchini}, {Manara}, {Schneider}, {Clarke},
  {Bouvier}, {Rosotti}, {Booth}, \& {Haworth}}]{facchini16}
{Facchini}, S., {Manara}, C.~F., {Schneider}, P.~C., {et~al.} 2016, \aap, 596,
  A38

\bibitem[{{Fang} {et~al.}(2013){Fang}, {Kim}, {van Boekel}, {Sicilia-Aguilar},
  {Henning}, \& {Flaherty}}]{Fang2013}
{Fang}, M., {Kim}, J.~S., {van Boekel}, R., {et~al.} 2013, \apjs, 207, 5

\bibitem[{{Feiden}(2016)}]{Feiden2016}
{Feiden}, G.~A. 2016, \aap, 593, A99

\bibitem[{{Fischer} {et~al.}(2022){Fischer}, {Hillenbrand}, {Herczeg},
  {Johnstone}, {K{\'o}sp{\'a}l}, \& {Dunham}}]{Fischer22}
{Fischer}, W.~J., {Hillenbrand}, L.~A., {Herczeg}, G.~J., {et~al.} 2022, arXiv
  e-prints, arXiv:2203.11257

\bibitem[{{Frasca} {et~al.}(2017){Frasca}, {Biazzo}, {Alcal{\'a}}, {Manara},
  {Stelzer}, {Covino}, \& {Antoniucci}}]{frasca2017}
{Frasca}, A., {Biazzo}, K., {Alcal{\'a}}, J.~M., {et~al.} 2017, \aap, 602, A33

\bibitem[{{Frasca} {et~al.}(2015){Frasca}, {Biazzo}, {Lanzafame}, {Alcal{\'a}},
  {Brugaletta}, {Klutsch}, {Stelzer}, {Sacco}, {Spina}, {Jeffries}, {Montes},
  {Alfaro}, {Barentsen}, {Bonito}, {Gameiro}, {L{\'o}pez-Santiago}, {Pace},
  {Pasquini}, {Prisinzano}, {Sousa}, {Gilmore}, {Randich}, {Micela},
  {Bragaglia}, {Flaccomio}, {Bayo}, {Costado}, {Franciosini}, {Hill},
  {Hourihane}, {Jofr{\'e}}, {Lardo}, {Maiorca}, {Masseron}, {Morbidelli}, \&
  {Worley}}]{frasca15}
{Frasca}, A., {Biazzo}, K., {Lanzafame}, A.~C., {et~al.} 2015, \aap, 575, A4

\bibitem[{{Gaia Collaboration} {et~al.}(2021){Gaia Collaboration}, {Brown},
  {Vallenari}, {Prusti}, {de Bruijne}, {Babusiaux}, {Biermann}, {Creevey},
  {Evans}, {Eyer}, {Hutton}, {Jansen}, {Jordi}, {Klioner}, {Lammers},
  {Lindegren}, {Luri}, {Mignard}, {Panem}, {Pourbaix}, {Randich}, {Sartoretti},
  {Soubiran}, {Walton}, {Arenou}, {Bailer-Jones}, {Bastian}, {Cropper},
  {Drimmel}, {Katz}, {Lattanzi}, {van Leeuwen}, {Bakker}, {Cacciari},
  {Casta{\~n}eda}, {De Angeli}, {Ducourant}, {Fabricius}, {Fouesneau},
  {Fr{\'e}mat}, {Guerra}, {Guerrier}, {Guiraud}, {Jean-Antoine Piccolo},
  {Masana}, {Messineo}, {Mowlavi}, {Nicolas}, {Nienartowicz}, {Pailler},
  {Panuzzo}, {Riclet}, {Roux}, {Seabroke}, {Sordo}, {Tanga}, {Th{\'e}venin},
  {Gracia-Abril}, {Portell}, {Teyssier}, {Altmann}, {Andrae}, {Bellas-Velidis},
  {Benson}, {Berthier}, {Blomme}, {Brugaletta}, {Burgess}, {Busso}, {Carry},
  {Cellino}, {Cheek}, {Clementini}, {Damerdji}, {Davidson}, {Delchambre},
  {Dell'Oro}, {Fern{\'a}ndez-Hern{\'a}ndez}, {Galluccio}, {Garc{\'\i}a-Lario},
  {Garcia-Reinaldos}, {Gonz{\'a}lez-N{\'u}{\~n}ez}, {Gosset}, {Haigron},
  {Halbwachs}, {Hambly}, {Harrison}, {Hatzidimitriou}, {Heiter},
  {Hern{\'a}ndez}, {Hestroffer}, {Hodgkin}, {Holl}, {Jan{\ss}en}, {Jevardat de
  Fombelle}, {Jordan}, {Krone-Martins}, {Lanzafame}, {L{\"o}ffler}, {Lorca},
  {Manteiga}, {Marchal}, {Marrese}, {Moitinho}, {Mora}, {Muinonen}, {Osborne},
  {Pancino}, {Pauwels}, {Petit}, {Recio-Blanco}, {Richards}, {Riello},
  {Rimoldini}, {Robin}, {Roegiers}, {Rybizki}, {Sarro}, {Siopis}, {Smith},
  {Sozzetti}, {Ulla}, {Utrilla}, {van Leeuwen}, {van Reeven}, {Abbas}, {Abreu
  Aramburu}, {Accart}, {Aerts}, {Aguado}, {Ajaj}, {Altavilla}, {{\'A}lvarez},
  {{\'A}lvarez Cid-Fuentes}, {Alves}, {Anderson}, {Anglada Varela}, {Antoja},
  {Audard}, {Baines}, {Baker}, {Balaguer-N{\'u}{\~n}ez}, {Balbinot}, {Balog},
  {Barache}, {Barbato}, {Barros}, {Barstow}, {Bartolom{\'e}}, {Bassilana},
  {Bauchet}, {Baudesson-Stella}, {Becciani}, {Bellazzini}, {Bernet}, {Bertone},
  {Bianchi}, {Blanco-Cuaresma}, {Boch}, {Bombrun}, {Bossini}, {Bouquillon},
  {Bragaglia}, {Bramante}, {Breedt}, {Bressan}, {Brouillet}, {Bucciarelli},
  {Burlacu}, {Busonero}, {Butkevich}, {Buzzi}, {Caffau}, {Cancelliere},
  {C{\'a}novas}, {Cantat-Gaudin}, {Carballo}, {Carlucci}, {Carnerero},
  {Carrasco}, {Casamiquela}, {Castellani}, {Castro-Ginard}, {Castro Sampol},
  {Chaoul}, {Charlot}, {Chemin}, {Chiavassa}, {Cioni}, {Comoretto}, {Cooper},
  {Cornez}, {Cowell}, {Crifo}, {Crosta}, {Crowley}, {Dafonte}, {Dapergolas},
  {David}, {David}, {de Laverny}, {De Luise}, {De March}, {De Ridder}, {de
  Souza}, {de Teodoro}, {de Torres}, {del Peloso}, {del Pozo}, {Delbo},
  {Delgado}, {Delgado}, {Delisle}, {Di Matteo}, {Diakite}, {Diener},
  {Distefano}, {Dolding}, {Eappachen}, {Edvardsson}, {Enke}, {Esquej}, {Fabre},
  {Fabrizio}, {Faigler}, {Fedorets}, {Fernique}, {Fienga}, {Figueras},
  {Fouron}, {Fragkoudi}, {Fraile}, {Franke}, {Gai}, {Garabato},
  {Garcia-Gutierrez}, {Garc{\'\i}a-Torres}, {Garofalo}, {Gavras}, {Gerlach},
  {Geyer}, {Giacobbe}, {Gilmore}, {Girona}, {Giuffrida}, {Gomel}, {Gomez},
  {Gonzalez-Santamaria}, {Gonz{\'a}lez-Vidal}, {Granvik},
  {Guti{\'e}rrez-S{\'a}nchez}, {Guy}, {Hauser}, {Haywood}, {Helmi}, {Hidalgo},
  {Hilger}, {H{\l}adczuk}, {Hobbs}, {Holland}, {Huckle}, {Jasniewicz},
  {Jonker}, {Juaristi Campillo}, {Julbe}, {Karbevska}, {Kervella}, {Khanna},
  {Kochoska}, {Kontizas}, {Kordopatis}, {Korn}, {Kostrzewa-Rutkowska},
  {Kruszy{\'n}ska}, {Lambert}, {Lanza}, {Lasne}, {Le Campion}, {Le Fustec},
  {Lebreton}, {Lebzelter}, {Leccia}, {Leclerc}, {Lecoeur-Taibi}, {Liao},
  {Licata}, {Lindstr{\o}m}, {Lister}, {Livanou}, {Lobel}, {Madrero Pardo},
  {Managau}, {Mann}, {Marchant}, {Marconi}, {Marcos Santos}, {Marinoni},
  {Marocco}, {Marshall}, {Martin Polo}, {Mart{\'\i}n-Fleitas}, {Masip},
  {Massari}, {Mastrobuono-Battisti}, {Mazeh}, {McMillan}, {Messina},
  {Michalik}, {Millar}, {Mints}, {Molina}, {Molinaro}, {Moln{\'a}r},
  {Montegriffo}, {Mor}, {Morbidelli}, {Morel}, {Morris}, {Mulone}, {Munoz},
  {Muraveva}, {Murphy}, {Musella}, {Noval}, {Ord{\'e}novic}, {Orr{\`u}},
  {Osinde}, {Pagani}, {Pagano}, {Palaversa}, {Palicio}, {Panahi}, {Pawlak},
  {Pe{\~n}alosa Esteller}, {Penttil{\"a}}, {Piersimoni}, {Pineau}, {Plachy},
  {Plum}, {Poggio}, {Poretti}, {Poujoulet}, {Pr{\v{s}}a}, {Pulone}, {Racero},
  {Ragaini}, {Rainer}, {Raiteri}, {Rambaux}, {Ramos}, {Ramos-Lerate}, {Re
  Fiorentin}, {Regibo}, {Reyl{\'e}}, {Ripepi}, {Riva}, {Rixon}, {Robichon},
  {Robin}, {Roelens}, {Rohrbasser}, {Romero-G{\'o}mez}, {Rowell}, {Royer},
  {Rybicki}, {Sadowski}, {Sagrist{\`a} Sell{\'e}s}, {Sahlmann}, {Salgado},
  {Salguero}, {Samaras}, {Sanchez Gimenez}, {Sanna}, {Santove{\~n}a},
  {Sarasso}, {Schultheis}, {Sciacca}, {Segol}, {Segovia}, {S{\'e}gransan},
  {Semeux}, {Shahaf}, {Siddiqui}, {Siebert}, {Siltala}, {Slezak}, {Smart},
  {Solano}, {Solitro}, {Souami}, {Souchay}, {Spagna}, {Spoto}, {Steele},
  {Steidelm{\"u}ller}, {Stephenson}, {S{\"u}veges}, {Szabados}, {Szegedi-Elek},
  {Taris}, {Tauran}, {Taylor}, {Teixeira}, {Thuillot}, {Tonello}, {Torra},
  {Torra}, {Turon}, {Unger}, {Vaillant}, {van Dillen}, {Vanel}, {Vecchiato},
  {Viala}, {Vicente}, {Voutsinas}, {Weiler}, {Wevers}, {Wyrzykowski}, {Yoldas},
  {Yvard}, {Zhao}, {Zorec}, {Zucker}, {Zurbach}, \& {Zwitter}}]{Gaia2021}
{Gaia Collaboration}, {Brown}, A.~G.~A., {Vallenari}, A., {et~al.} 2021, \aap,
  649, A1

\bibitem[{{Hartmann} {et~al.}(2016){Hartmann}, {Herczeg}, \&
  {Calvet}}]{hartmann16}
{Hartmann}, L., {Herczeg}, G., \& {Calvet}, N. 2016, \araa, 54, 135

\bibitem[{{Herczeg} \& {Hillenbrand}(2014)}]{HH14}
{Herczeg}, G.~J. \& {Hillenbrand}, L.~A. 2014, \apj, 786, 97

\bibitem[{{Hillenbrand} \& {Findeisen}(2015)}]{hillenbrand15}
{Hillenbrand}, L.~A. \& {Findeisen}, K.~P. 2015, \apj, 808, 68

\bibitem[{{Kafka}(2020)}]{Kafka2020}
{Kafka}, S. 2020, in European Planetary Science Congress, EPSC2020--314

\bibitem[{{K{\'o}sp{\'a}l}(2011)}]{kospal2011FUor}
{K{\'o}sp{\'a}l}, {\'A}. 2011, \aap, 535, A125

\bibitem[{{K{\'o}sp{\'a}l} {et~al.}(2012){K{\'o}sp{\'a}l}, {{\'A}brah{\'a}m},
  {Acosta-Pulido}, {Dullemond}, {Henning}, {Kun}, {Leinert}, {Mo{\'o}r}, \&
  {Turner}}]{kospal12}
{K{\'o}sp{\'a}l}, {\'A}., {{\'A}brah{\'a}m}, P., {Acosta-Pulido}, J.~A.,
  {et~al.} 2012, \apjs, 201, 11

\bibitem[{{K{\'o}sp{\'a}l} {et~al.}(2011){K{\'o}sp{\'a}l}, {{\'A}brah{\'a}m},
  {Goto}, {Reg{\'a}ly}, {Dullemond}, {Henning}, {Juh{\'a}sz},
  {Sicilia-Aguilar}, \& {van den Ancker}}]{kospal11EXor}
{K{\'o}sp{\'a}l}, {\'A}., {{\'A}brah{\'a}m}, P., {Goto}, M., {et~al.} 2011,
  \apj, 736, 72

\bibitem[{{Koutoulaki} {et~al.}(2019){Koutoulaki}, {Facchini}, {Manara},
  {Natta}, {Garcia Lopez}, {Fedriani}, {Caratti o Garatti}, {Coffey}, \&
  {Ray}}]{koutoulaki19}
{Koutoulaki}, M., {Facchini}, S., {Manara}, C.~F., {et~al.} 2019, \aap, 625,
  A49

\bibitem[{{Kraus} \& {Hillenbrand}(2007)}]{Kraus2007}
{Kraus}, A.~L. \& {Hillenbrand}, L.~A. 2007, \apj, 662, 413

\bibitem[{{Lodato} {et~al.}(2017){Lodato}, {Scardoni}, {Manara}, \&
  {Testi}}]{lodato17}
{Lodato}, G., {Scardoni}, C.~E., {Manara}, C.~F., \& {Testi}, L. 2017, \mnras,
  472, 4700

\bibitem[{{Long} {et~al.}(2017){Long}, {Herczeg}, {Pascucci}, {Drabek-Maunder},
  {Mohanty}, {Testi}, {Apai}, {Hendler}, {Henning}, {Manara}, \&
  {Mulders}}]{long17}
{Long}, F., {Herczeg}, G.~J., {Pascucci}, I., {et~al.} 2017, \apj, 844, 99

\bibitem[{{Manara} {et~al.}(2022){Manara}, {Ansdell}, {Rosotti}, {Hughes},
  {Armitage}, {Lodato}, \& {Williams}}]{manara22}
{Manara}, C.~F., {Ansdell}, M., {Rosotti}, G.~P., {et~al.} 2022, arXiv
  e-prints, arXiv:2203.09930

\bibitem[{{Manara} {et~al.}(2013){Manara}, {Beccari}, {Da Rio}, {De Marchi},
  {Natta}, {Ricci}, {Robberto}, \& {Testi}}]{manara13b}
{Manara}, C.~F., {Beccari}, G., {Da Rio}, N., {et~al.} 2013, \aap, 558, A114

\bibitem[{{Manara} {et~al.}(2016){Manara}, {Fedele}, {Herczeg}, \&
  {Teixeira}}]{manara16a}
{Manara}, C.~F., {Fedele}, D., {Herczeg}, G.~J., \& {Teixeira}, P.~S. 2016,
  \aap, 585, A136

\bibitem[{{Manara} {et~al.}(2021){Manara}, {Frasca}, {Venuti}, {Siwak},
  {Herczeg}, {Calvet}, {Hernandez}, {Tychoniec}, {Gangi}, {Alcal{\'a}},
  {Boffin}, {Nisini}, {Robberto}, {Briceno}, {Campbell-White},
  {Sicilia-Aguilar}, {McGinnis}, {Fedele}, {K{\'o}sp{\'a}l}, {{\'A}brah{\'a}m},
  {Alonso-Santiago}, {Antoniucci}, {Arulanantham}, {Bacciotti}, {Banzatti},
  {Beccari}, {Benisty}, {Biazzo}, {Bouvier}, {Cabrit}, {Caratti o Garatti},
  {Coffey}, {Covino}, {Dougados}, {Eisl{\"o}ffel}, {Ercolano}, {Espaillat},
  {Erkal}, {Facchini}, {Fang}, {Fiorellino}, {Fischer}, {France}, {Gameiro},
  {Garcia Lopez}, {Giannini}, {Ginski}, {Grankin}, {G{\"u}nther}, {Hartmann},
  {Hillenbrand}, {Hussain}, {James}, {Koutoulaki}, {Lodato}, {Mauc{\'o}},
  {Mendigut{\'\i}a}, {Mentel}, {Miotello}, {Oudmaijer}, {Rigliaco}, {Rosotti},
  {Sanchis}, {Schneider}, {Spina}, {Stelzer}, {Testi}, {Thanathibodee}, {Vink},
  {Walter}, {Williams}, \& {Zsidi}}]{manara21}
{Manara}, C.~F., {Frasca}, A., {Venuti}, L., {et~al.} 2021, \aap, 650, A196

\bibitem[{{Manara} {et~al.}(2020){Manara}, {Natta}, {Rosotti}, {Alcal{\'a}},
  {Nisini}, {Lodato}, {Testi}, {Pascucci}, {Hillenbrand}, {Carpenter},
  {Scholz}, {Fedele}, {Frasca}, {Mulders}, {Rigliaco}, {Scardoni}, \&
  {Zari}}]{manara20}
{Manara}, C.~F., {Natta}, A., {Rosotti}, G.~P., {et~al.} 2020, \aap, 639, A58

\bibitem[{{Manara} {et~al.}(2017){Manara}, {Testi}, {Herczeg}, {Pascucci},
  {Alcal{\'a}}, {Natta}, {Antoniucci}, {Fedele}, {Mulders}, {Henning},
  {Mohanty}, {Prusti}, \& {Rigliaco}}]{manara17a}
{Manara}, C.~F., {Testi}, L., {Herczeg}, G.~J., {et~al.} 2017, \aap, 604, A127

\bibitem[{{Modigliani} {et~al.}(2010){Modigliani}, {Goldoni}, {Royer},
  {Haigron}, {Guglielmi}, {Fran{\c{c}}ois}, {Horrobin}, {Bristow}, {Vernet},
  {Moehler}, {Kerber}, {Ballester}, {Mason}, \& {Christensen}}]{xspipe}
{Modigliani}, A., {Goldoni}, P., {Royer}, F., {et~al.} 2010, in Society of
  Photo-Optical Instrumentation Engineers (SPIE) Conference Series, Vol. 7737,
  Observatory Operations: Strategies, Processes, and Systems III, ed. D.~R.
  {Silva}, A.~B. {Peck}, \& B.~T. {Soifer}, 773728

\bibitem[{{Mulders} {et~al.}(2017){Mulders}, {Pascucci}, {Manara}, {Testi},
  {Herczeg}, {Henning}, {Mohanty}, \& {Lodato}}]{mulders17}
{Mulders}, G.~D., {Pascucci}, I., {Manara}, C.~F., {et~al.} 2017, \apj, 847, 31

\bibitem[{{Pascucci} {et~al.}(2016){Pascucci}, {Testi}, {Herczeg}, {Long},
  {Manara}, {Hendler}, {Mulders}, {Krijt}, {Ciesla}, {Henning}, {Mohanty},
  {Drabek-Maunder}, {Apai}, {Sz{\H u}cs}, {Sacco}, \& {Olofsson}}]{pascucci16}
{Pascucci}, I., {Testi}, L., {Herczeg}, G.~J., {et~al.} 2016, \apj, 831, 125

\bibitem[{{Roman-Duval} {et~al.}(2020){Roman-Duval}, {Proffitt}, {Taylor},
  {Monroe}, {Fischer}, {Fischer}, {Fullerton}, {Aloisi}, {Britt}, {Busko},
  {Carlberg}, {De Rosa}, {Jedrzejewski}, {Lockwood}, {Frazer}, {Hernandez},
  {James}, {Oliveira}, {Plesha}, {Riedel}, {Riley}, {Sahnow}, {Sankrit},
  {Shaw}, {Smith}, {Sohn}, {Som}, {Ubeda}, \& {Welty}}]{ullysesDR1}
{Roman-Duval}, J., {Proffitt}, C.~R., {Taylor}, J.~M., {et~al.} 2020, Research
  Notes of the American Astronomical Society, 4, 205

\bibitem[{{Rosotti} {et~al.}(2017){Rosotti}, {Clarke}, {Manara}, \&
  {Facchini}}]{rosotti17}
{Rosotti}, G.~P., {Clarke}, C.~J., {Manara}, C.~F., \& {Facchini}, S. 2017,
  \mnras, 468, 1631

\bibitem[{{Rugel} {et~al.}(2018){Rugel}, {Fedele}, \& {Herczeg}}]{Rugel2018}
{Rugel}, M., {Fedele}, D., \& {Herczeg}, G. 2018, \aap, 609, A70

\bibitem[{{Schisano} {et~al.}(2009){Schisano}, {Covino}, {Alcal{\'a}},
  {Esposito}, {Gandolfi}, \& {Guenther}}]{Schisano2009}
{Schisano}, E., {Covino}, E., {Alcal{\'a}}, J.~M., {et~al.} 2009, \aap, 501,
  1013

\bibitem[{{Schneider} {et~al.}(2015){Schneider}, {France}, {G{\"u}nther},
  {Herczeg}, {Robrade}, {Bouvier}, {McJunkin}, \& {Schmitt}}]{Schneider2015}
{Schneider}, P.~C., {France}, K., {G{\"u}nther}, H.~M., {et~al.} 2015, \aap,
  584, A51

\bibitem[{{Schneider} {et~al.}(2018){Schneider}, {Manara}, {Facchini},
  {G{\"u}nther}, {Herczeg}, {Fedele}, \& {Teixeira}}]{schneider18}
{Schneider}, P.~C., {Manara}, C.~F., {Facchini}, S., {et~al.} 2018, \aap, 614,
  A108

\bibitem[{{Smette} {et~al.}(2015){Smette}, {Sana}, {Noll}, {Horst}, {Kausch},
  {Kimeswenger}, {Barden}, {Szyszka}, {Jones}, {Gallenne}, {Vinther},
  {Ballester}, \& {Taylor}}]{molecfit1}
{Smette}, A., {Sana}, H., {Noll}, S., {et~al.} 2015, \aap, 576, A77

\bibitem[{{Stauffer} {et~al.}(2014){Stauffer}, {Cody}, {Baglin}, {Alencar},
  {Rebull}, {Hillenbrand}, {Venuti}, {Turner}, {Carpenter}, {Plavchan},
  {Findeisen}, {Carey}, {Terebey}, {Morales-Calder{\'o}n}, {Bouvier}, {Micela},
  {Flaccomio}, {Song}, {Gutermuth}, {Hartmann}, {Calvet}, {Whitney}, {Barrado},
  {Vrba}, {Covey}, {Herbst}, {Furesz}, {Aigrain}, \& {Favata}}]{stauffer14}
{Stauffer}, J., {Cody}, A.~M., {Baglin}, A., {et~al.} 2014, \aj, 147, 83

\bibitem[{{Tabone} {et~al.}(2022){Tabone}, {Rosotti}, {Lodato}, {Armitage},
  {Cridland}, \& {van Dishoeck}}]{tabone21}
{Tabone}, B., {Rosotti}, G.~P., {Lodato}, G., {et~al.} 2022, \mnras, 512, L74

\bibitem[{{Venuti} {et~al.}(2014){Venuti}, {Bouvier}, {Flaccomio}, {Alencar},
  {Irwin}, {Stauffer}, {Cody}, {Teixeira}, {Sousa}, {Micela}, {Cuillandre}, \&
  {Peres}}]{venuti14}
{Venuti}, L., {Bouvier}, J., {Flaccomio}, E., {et~al.} 2014, \aap, 570, A82

\bibitem[{{Venuti} {et~al.}(2015){Venuti}, {Bouvier}, {Irwin}, {Stauffer},
  {Hillenbrand}, {Rebull}, {Cody}, {Alencar}, {Micela}, {Flaccomio}, \&
  {Peres}}]{venuti15}
{Venuti}, L., {Bouvier}, J., {Irwin}, J., {et~al.} 2015, \aap, 581, A66

\bibitem[{{Vernet} {et~al.}(2011){Vernet}, {Dekker}, {D'Odorico}, {Kaper},
  {Kjaergaard}, {Hammer}, {Randich}, {Zerbi}, {Groot}, {Hjorth}, {Guinouard},
  {Navarro}, {Adolfse}, {Albers}, {Amans}, {Andersen}, {Andersen}, {Binetruy},
  {Bristow}, {Castillo}, {Chemla}, {Christensen}, {Conconi}, {Conzelmann},
  {Dam}, {de Caprio}, {de Ugarte Postigo}, {Delabre}, {di Marcantonio},
  {Downing}, {Elswijk}, {Finger}, {Fischer}, {Flores}, {Fran{\c{c}}ois},
  {Goldoni}, {Guglielmi}, {Haigron}, {Hanenburg}, {Hendriks}, {Horrobin},
  {Horville}, {Jessen}, {Kerber}, {Kern}, {Kiekebusch}, {Kleszcz}, {Klougart},
  {Kragt}, {Larsen}, {Lizon}, {Lucuix}, {Mainieri}, {Manuputy}, {Martayan},
  {Mason}, {Mazzoleni}, {Michaelsen}, {Modigliani}, {Moehler}, {M{\o}ller},
  {Norup S{\o}rensen}, {N{\o}rregaard}, {P{\'e}roux}, {Patat}, {Pena}, {Pragt},
  {Reinero}, {Rigal}, {Riva}, {Roelfsema}, {Royer}, {Sacco}, {Santin},
  {Schoenmaker}, {Spano}, {Sweers}, {Ter Horst}, {Tintori}, {Tromp}, {van
  Dael}, {van der Vliet}, {Venema}, {Vidali}, {Vinther}, {Vola}, {Winters},
  {Wistisen}, {Wulterkens}, \& {Zacchei}}]{vernet11}
{Vernet}, J., {Dekker}, H., {D'Odorico}, S., {et~al.} 2011, \aap, 536, A105

\bibitem[{{Zsidi} {et~al.}(2022){Zsidi}, {Manara}, {K{\'o}sp{\'a}l}, {Hussain},
  {{\'A}brah{\'a}m}, {Alecian}, {B{\'o}di}, {P{\'a}l}, \& {Sarkis}}]{zsidi22}
{Zsidi}, G., {Manara}, C.~F., {K{\'o}sp{\'a}l}, {\'A}., {et~al.} 2022, arXiv
  e-prints, arXiv:2201.03396

\end{thebibliography}

\appendix

\section{Broadband variability}

\subsection{Photometric variability}\label{photometry}
\begin{figure*}
     \centering
     \includegraphics[width = \textwidth]{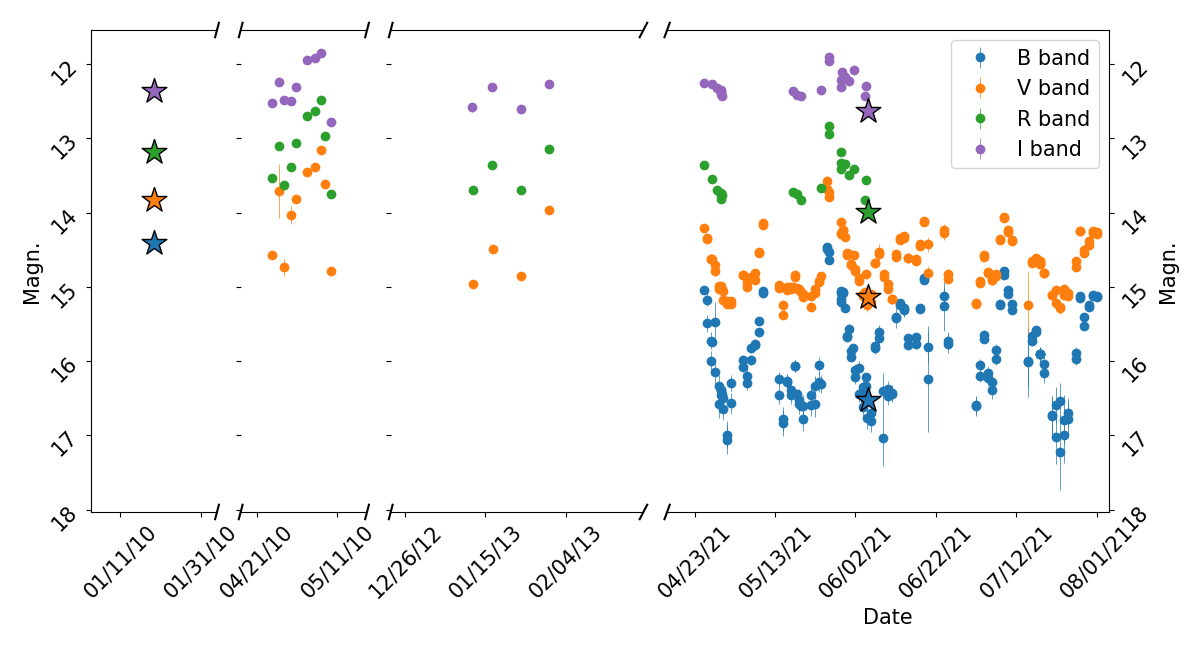}
     \caption{Comparison between the synthetic photometry and the available photometric data. The stars indicate synthetic photometry measured on the X-Shooter data in 2010 and 2021. The colors correspond to the same band as the photometric data.}
     \label{fig:synthPhot}
\end{figure*}

The available photometric data presented in Sect.~\ref{sec:phot_data} is shown in Fig.~\ref{SynthPhot} as a function of the time of the observations. On the same plot, the synthetic photometry derived from the X-Shooter spectra (see Sect.~\ref{sec:cont} for the procedure and Table~\ref{SynthPhot} for the values) is also included for comparison. 
The synthetic photometry measured on the 2021 spectrum agrees with the photometry taken close in time. When looking at the light curve around the 2021 observations, the brightness of the target at the time of the 2021 X-Shooter observations is close to the local variability minimum. Indeed, there seems to be a quasi-periodic variability pattern with timescales of $\sim$22 days which, however, is only hinted at in the periodogram analysis we carried out, possibly due to the short time coverage with respect to the length of this putative period. 
The 2010 synthetic data fall on the high side, but within the range of variability for all epochs of observation.

\begin{figure}
    \centering
    \includegraphics[width = 0.5\textwidth]{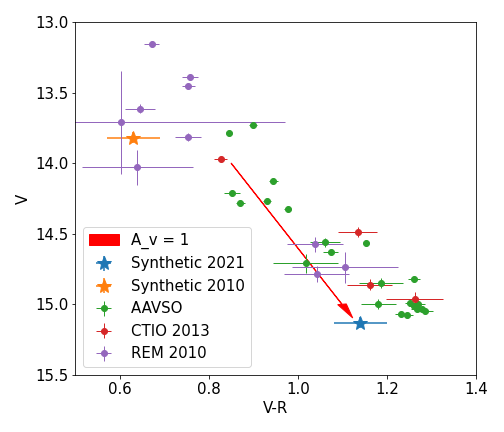}
    \caption{$V-R$ color-magnitude diagram for the AAVSO photometric bands. The blue dots show the AAVSO photometry. The color was computed by combining the closest in time observations during the same nights. The red arrow displays the reddening vector for $A_V = 1$ mag and $R_V = 3.1$ using the extinction law by \citet{cardelli98}. Synthetic photometry data obtained on the X-Shooter spectra and REM photometry is also shown.}
    \label{fig:VVSv-r}
\end{figure}

\begin{figure}
    \centering
    \includegraphics[width = 0.5\textwidth]{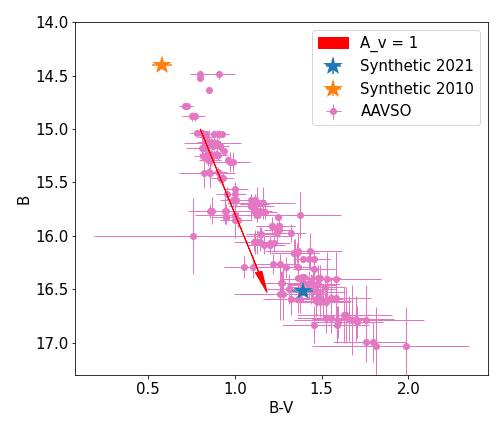}
    \caption{Same as Fig.~\ref{fig:VVSv-r}, but for the $B-V$ color. }
    \label{fig:bVSb-v}
\end{figure}

The variability amplitude observed in the photometry is larger for bluer bands. Such a trend is expected in the case of a variable accretion rate or reddening. To further explore this effect, we show in Fig.~\ref{fig:VVSv-r} the $V$ versus $V-R$ and in Fig. \ref{fig:bVSb-v} the $B$ versus $B-V$ color-magnitude diagram (CMD), respectively, and a reddening vector with $R_V$=3.1. The reddening vector is well aligned with the AAVSO data in the $V-R$ CMD, and less so in the $B-V$ one. In either case, the reddening vector cannot reproduce the variability between the 2010 and 2021 spectra analyzed here. The variability, or at least a part of it, must be ascribed to accretion variability.

\begin{table}[t]
\centering
\caption{Synthetic photometry on the X-Shooter spectra}
\label{SynthPhot}
\begin{tabular}{c|cc}
\hline \hline 
Band & 2010 & 2021 \\ \hline 
$B$ & 14.40 &   16.52   \\
$V$ & 13.82 &  15.13   \\
$R$ & 13.19 &   13.99   \\
$I$ & 12.36  &    12.64  \\
 \hline
\end{tabular}
\end{table}

\subsection{Continuum variability}\label{appendix:Red}
\begin{figure}
    \centering
    \includegraphics[width = 0.499\textwidth]{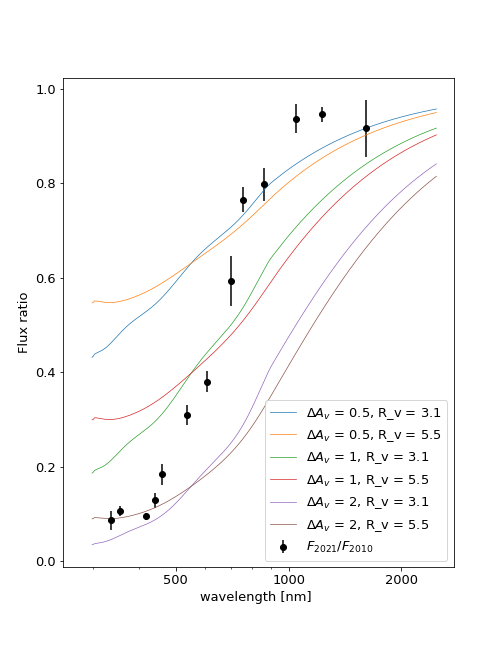}
    \caption{Comparison of the flux ratio of the observed spectra with the ratio expected from a difference in reddening $\Delta A_V$ applied to the two spectra. The black points indicate the average ratio of the observed spectra in wavelength windows of 4nm. The colored lines indicate the expected ratio from the Cardelli reddening law \citep{cardelli98}. 
    }
    \label{Fig:FluxRatioVSred}
\end{figure}

We computed the flux ratio between the 2010 and 2021 spectra as $F_{2021}/F_{2010}$ using the median flux in an interval of 4 nm around a number of wavelengths in the spectra, chosen to sample the continuum in the spectra and avoiding emission lines. The errors were then computed from the standard deviation of the ratios within these intervals. 
Fig. \ref{Fig:FluxRatioVSred} displays the measured flux ratio of the two spectra along with the ratio expected from a difference in extinction ($\Delta A_V$) following the \citet{cardelli98} reddening law.

The measured flux ratio between the spectra is steeper than any reasonable extinction difference and reddening law.  This confirms that the difference in both spectra cannot be attributed to a variation in extinction between the two epochs, as described in Sect.~\ref{photometry}.

\clearpage

\section{Additional analysis of the spectra}
\FloatBarrier
\subsection{Variations in the lithium absorption}

The lithium absorption line at $\lambda=$ 670.78 nm is one of the strongest absorption lines in low-mass CTTS. The depth of this line solely depends on the stellar temperature and on the age of the target. For this reason, this is a key feature to constrain whether the variations seen in the spectra are due to accretion, which would result in a higher veiling of this line, or extinction, which would leave the line depth unaffected. 

We decided to compare the depth of this line, not only between the 2010 and 2021 epochs of the X-Shooter observations, but to also include the spectra of XX Cha taken with the UVES instrument on the VLT within $\pm$2 days from the 2021 observations with X-Shooter. The UVES data were reduced with the ESO Reflex pipeline as described in \citet{manara21}. 

\begin{figure}
    \centering
    \includegraphics[width=0.5\textwidth]{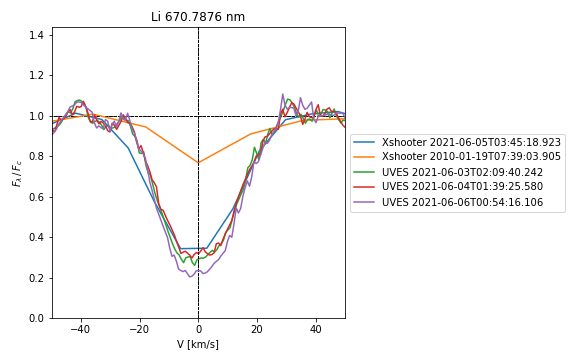}
     \caption{Lithium 670.78 nm line is shown normalized to the local continuum and as a function of velocity, after correcting for barycentric and radial velocity. The dates of the observations are reported in the legend.} 
    \label{fig:LiLine}
\end{figure}

The continuum normalized lithium absorption line is shown in Fig.~\ref{fig:LiLine}. The four spectra obtained in 2021 all show very similar depths, with differences mainly being due to the lower resolution of the X-Shooter observations with respect to the UVES ones. On the other hand, the spectrum obtained in 2010 shows a significantly shallower depth in this line, which is a clear signature of an increased veiling in this epoch. This should thus be ascribed to a stronger accretion rate in 2010.

\subsection{Fitting of the X-Shooter spectra}\label{App:fit}

To derive the stellar and accretion properties, we used the method by \citet{manara13b}. This method consists of fitting three components to the observed spectra: a Class III template to reproduce the stellar photosphere, a slab model for the continuum excess emission due to accretion, and the reddening law by \citet{cardelli98} to account for interstellar extinction. 
The slab model and Class III template were scaled to match the observed flux at $360$ and at $710$ nm. A $ \chi ^2_{\rm like}$ function was computed for a grid of parameters considering different extinction values, slab model parameters and normalization, and Class III templates and their normalizations. 
This $ \chi ^2_{\rm like}$ was computed in some key spectral features, such as the Balmer jump, Balmer continuum, Paschen continuum, and a number of molecular bands that are indicative of the spectral type of the target.

\begin{figure}
    \centering
    \includegraphics[width = 0.48\textwidth]{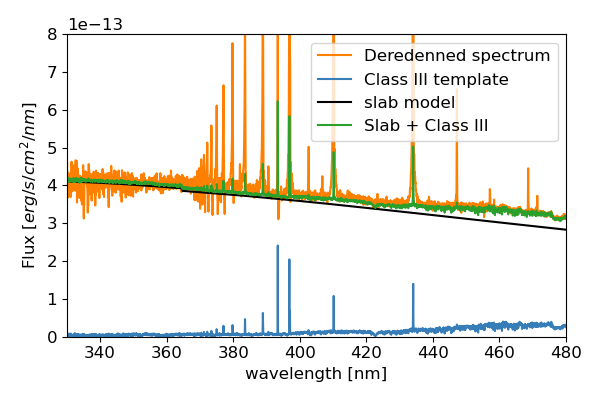}
    \caption{Best fit of the 2010 X-Shooter spectrum of XX Cha, with the best-fit parameters reported by \citet{manara16a}.}
    \label{fig:bestfit10balmer_av10}
\end{figure}

\begin{figure}
    \centering
    \includegraphics[width = 0.48\textwidth]{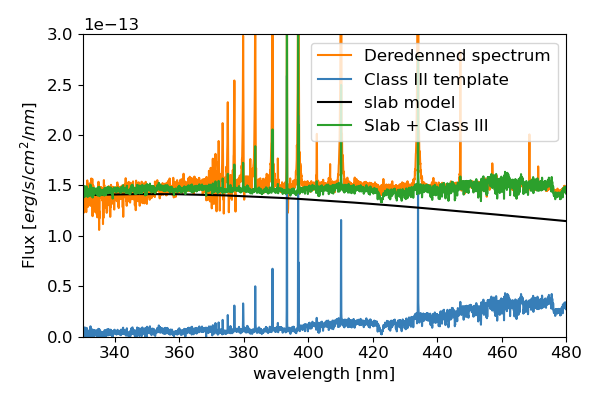}
    \caption{Best fit of the 2010 X-Shooter spectrum of XX Cha, assuming $A_V$ = 0.3 mag, referred to as 2010* in the text.}
    \label{fig:bestfit10balmer_av03}
\end{figure}

\begin{figure}
    \centering
    \includegraphics[width = 0.48\textwidth]{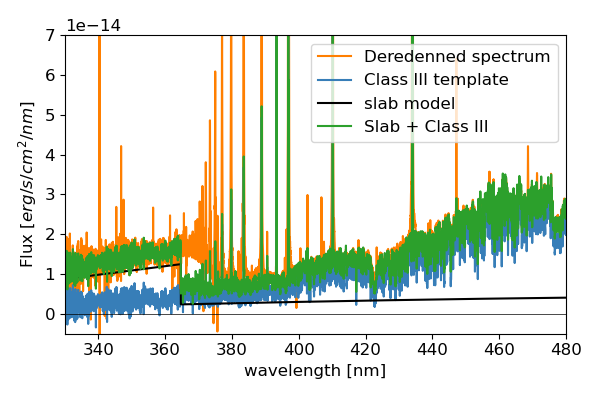}
    \caption{Best fit of the 2021 X-Shooter spectrum of XX Cha.}
    \label{fig:bestfit21balmer}
\end{figure}

Figures \ref{fig:bestfit10balmer_av10}, \ref{fig:bestfit10balmer_av03} and \ref{fig:bestfit21balmer} display the best-fit components around the Balmer jump. Comparing both figures highlights the significant differences in the accretion spectrum between the epochs.


\subsection{Accretion luminosity from emission lines}\label{Lines}

\begin{figure}
    \centering
    \includegraphics[width=0.5\textwidth]{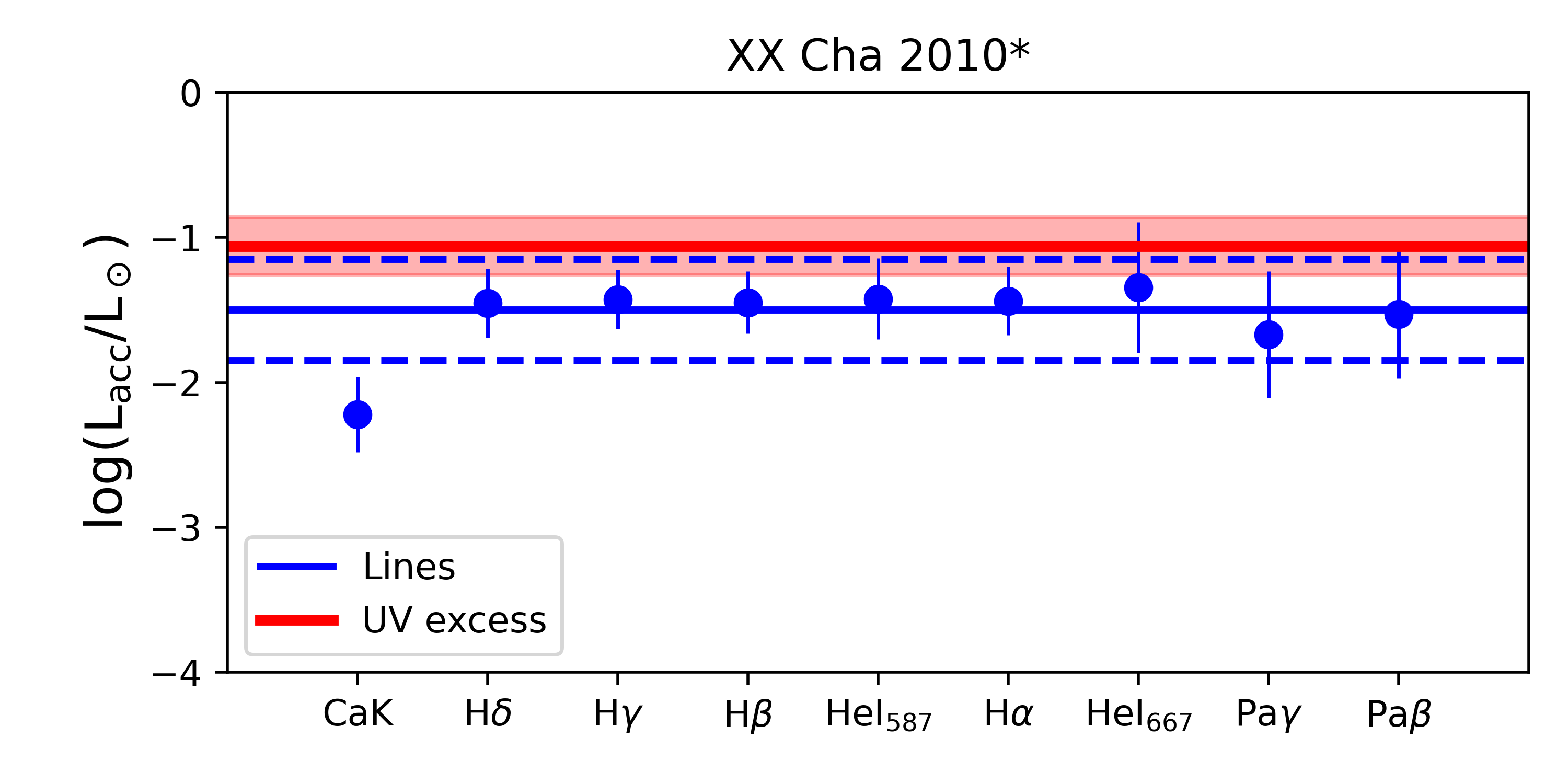}
    \caption{Comparison of the accretion luminosity derived from emission lines in the 2010* epoch with that derived from the UV excess.
    The mean accretion luminosity obtained from the lines is indicated with the solid blue line. The dashed lines indicates the error on the mean value. The accretion luminosity  obtained from the UV excess is displayed with the solid red line and its error is shown via the shaded area.}
    \label{fig:LaccLines2010*}
\end{figure}

\begin{figure}
    \centering
    \includegraphics[width=0.5\textwidth]{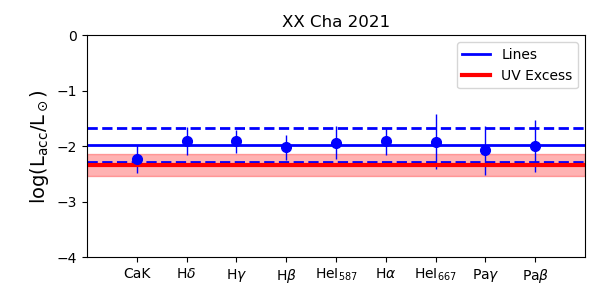}
    \caption{Comparison of the accretion luminosity derived from emission lines in the 2021 epoch with that derived from the UV excess.
    The mean accretion luminosity obtained from the lines is indicated with the solid blue line. The dashed lines indicates the error on the mean value. The accretion luminosity  obtained from the UV excess is displayed with the red solid line and its error is shown via the shaded area.}
    \label{fig:LaccLines2021}
\end{figure}

\begin{figure}
    \centering
    \includegraphics[width=0.5\textwidth]{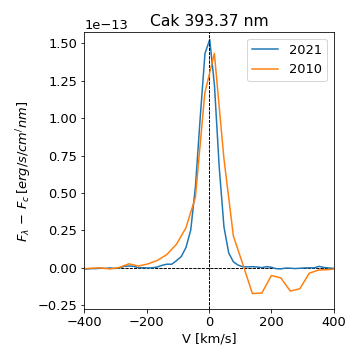}
\caption{Profile of the CaK 393.37 nm line in the two X-Shooter observations. }
    \label{fig:CaKLine}
\end{figure}

The accretion luminosities obtained from individual lines (see Sect.~\ref{sec:lline}) in both epochs are listed in table \ref{sec2:LineAccretion}.
These values are also plotted in Figs. \ref{fig:LaccLines2010*} and \ref{fig:LaccLines2021}, where they are compared to the results obtained from the UV excess in the corresponding epochs. The lines are ordered with increasing wavelengths, allowing one to discern possible trends with wavelength, which would indicate a wrong correction for extinction. Such a trend is not observed when using the value $A_V = 0.3$ mag for the 2010 epoch, whereas a slight dependence with wavelength is present at higher values of $A_V=1$ mag (see Fig. ~\ref{fig:LaccLines2010*}). 
The accretion luminosity obtained from the CaK line is lower than that found from the other lines in both epochs. This difference is more pronounced in 2010. The accretion luminosity derived from this line is even lower in 2010 than in 2021. 
In Fig. \ref{fig:CaKLine} we show the profile of this line in the two X-Shooter observations, and we observe an inverse P-Cygni profile in the 2010 observation, with an otherwise similar profile and intensity. The additional absorption seen here is a likely cause for the low accretion luminosity obtained. However, the Pa$\beta$, Pa$\gamma$, and HeI line at 667.82 nm also display a similar absorption feature in 2010, but the derived \lacc\, are larger than in 2021. 

\begin{figure*}
    \centering
    \includegraphics[width=0.49\textwidth]{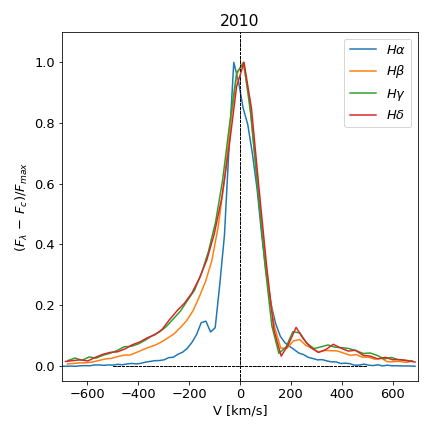}
    \includegraphics[width=0.49\textwidth]{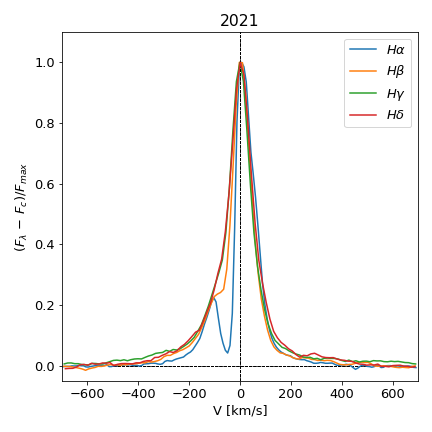}
    \caption{Comparison of the Balmer line profiles for the 2010 and 2021 X-Shooter spectra.}
    \label{fig:balmer}
\end{figure*}

The H$\beta$, H$\gamma$, and H$\delta$ lines profiles also change between both epochs. In 2021 they appear almost symmetrical, whereas in 2010 they are more extended on the blueshifted side. This is illustrated in figure \ref{fig:balmer}, and further discussed in Sect.~\ref{sec:lline}.

\begin{table}[t]

\centering
\caption{Accretion parameters derived from the emission lines.  }
\label{sec2:LineAccretion}
\begin{tabular}{c|c|c}
\hline \hline 
 Epoch of observation  
   & 2010*
           &2021\\
\hline
    & $\log(L_{\rm acc}/L_\odot)$  & $\log(L_{\rm acc}/L_\odot)$ \\
\hline
H$\alpha$ & -1.43 &    -1.91  \\
H$\beta$  & -1.45  & -2.02  \\
H$\gamma$ & -1.43 &   -1.91 \\
H$\delta$  & -1.45 &   -1.90  \\
CaK        & -2.22 & -2.23 \\
HeI$_{587}$ & -1.42 &  -1.93  \\
HeI$_{667}$ & -1.34  & -1.91  \\
Pa $\beta$& -1.52 &  -2.00 \\
Pa $\gamma$ & -1.66 &  -2.07  \\
mean & $-1.50 \pm  0.35$ & $-1.98 \pm 0.35$ \\
\hline
$\log$ (\macc /(\msun /yr))& $-8.1 \pm 0.4$ & $-8.7 \pm 0.4$ \\
\hline

 \hline
 \end{tabular}
 \tablefoot{The error on the mean $\log(L_{\rm acc}/L_\odot)$ is the standard deviation of the values obtained for the lines. }
\end{table}

\begin{figure}
    \centering
    \includegraphics[width=0.5\textwidth]{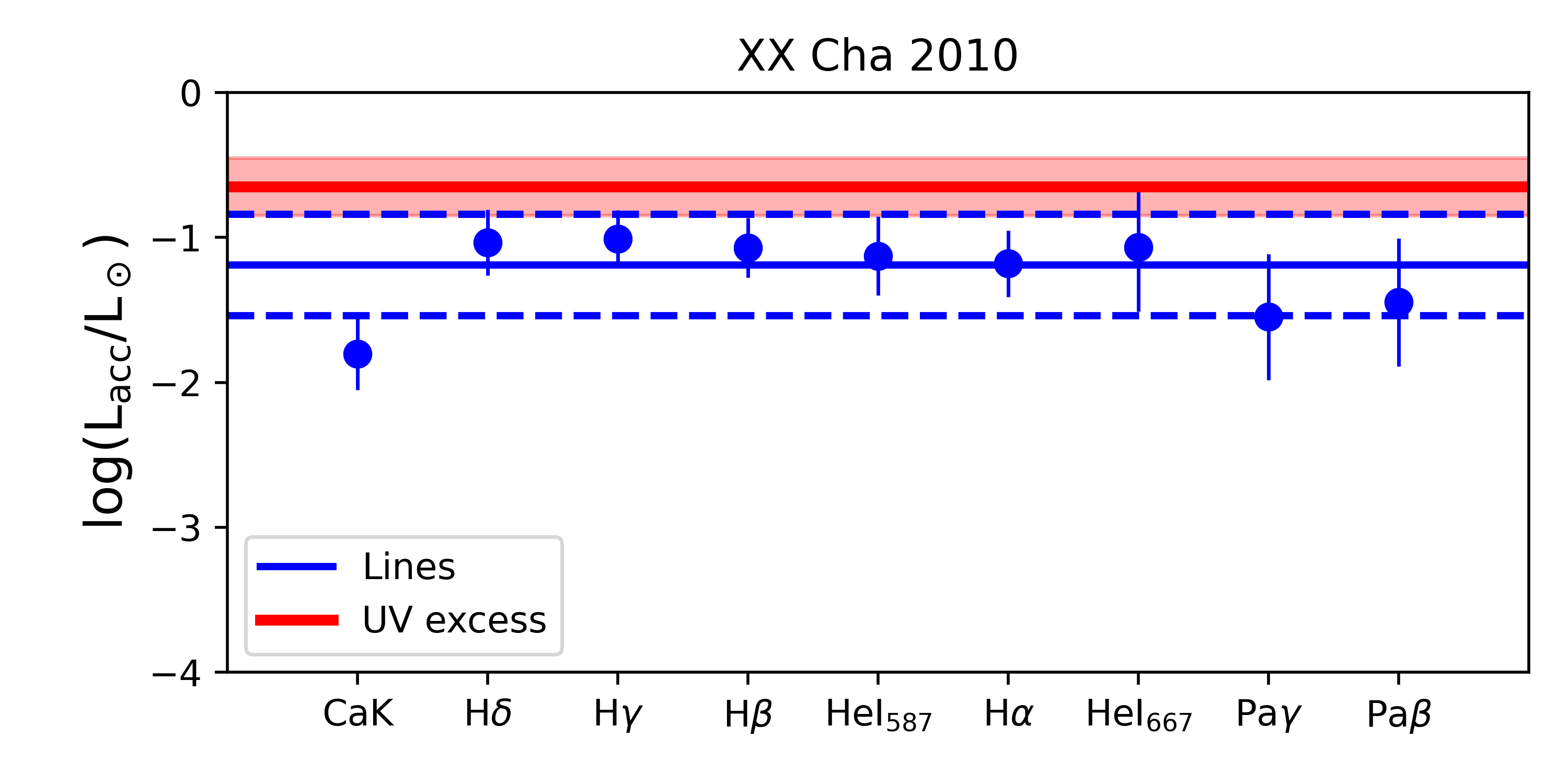}
    \caption{Comparison of the accretion luminosity derived from emission lines in the 2010 epoch with that derived from the UV excess.
    The mean accretion luminosity obtained from the lines is indicated with the solid blue line. The dashed lines indicates the error on the mean value. The accretion luminosity  obtained from the UV excess is displayed with the red solid line and its error is shown via the shaded area.}
    \label{fig:LaccLines2010}
\end{figure}

\FloatBarrier
\section{\macc -\mstar relation} 
\begin{figure}
   \centering
   \includegraphics[width=0.45\textwidth]{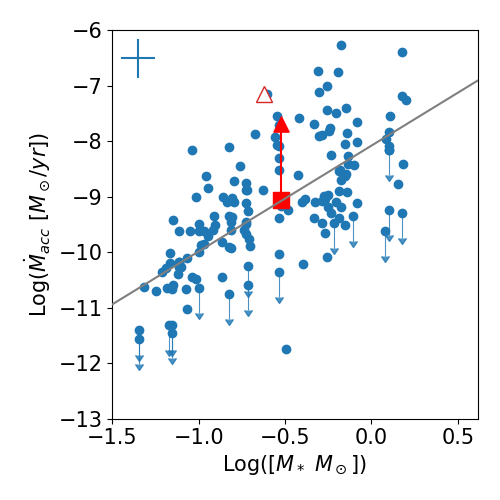}
      \caption{Measured \macc \, and \mstar\, for the targets in the Lupus and Chamaeleon~I star-forming regions observed with X-Shooter. The red-filled (empty) triangle and square indicate the 2010* (2010) and 2021 observations of XX Cha, respectively. The assumed errors for each data point is displayed on the top left-hand side. The gray line indicates the best-fit linear regression to the data.}
              
         \label{fig:macc_mstar}
   \end{figure}

Figure \ref{fig:macc_mstar} displays the measured \macc \, and \mstar\, for the targets in the Lupus and Chamaeleon~I star-forming regions observed with X-Shooter. The data are taken from \citet{manara22}, as in Fig.~\ref{fig::macc_mdisk}. 
XX Cha, reported with red symbols, is observed to be within the observed scatter in both epochs. 

\end{document}